# Exploring Cordial Labeling Techniques in Brain Connectivity Networks

**S. SOUNDAR RAJAN**[1] and **J. BASKAR BABUJEE**[2]

Department of Applied Sciences and Humanities
Madras Institute of Technology, Anna University, Chennai-44.

## ABSTRACT

*Graph-theoretical labeling provides a rigorous mathematical framework for characterizing the structural and functional organization of complex networks. This paper investigates the application of cordial labeling and signed product cordial labeling to brain connectivity graphs, emphasizing their relevance to small-world network models in neuroscience. The cordial condition is interpreted as a measure of structural balance between excitatory and inhibitory neuronal interactions, while the signed product cordial labeling reflects the coexistence of cooperative and antagonistic neural dynamics.*

## 1. INTRODUCTION

The human brain is an extraordinarily intricate system, consisting of approximately 86 billion neurons interconnected by nearly 150 trillion synapses. These synapses facilitate the transmission of electrical and chemical signals across neural circuits. In recent decades, modeling the brain as a complex network has become a central focus of neuroscience, enabling deeper insight into cognition, behavior, and perception. The study of brain

connectivity—whether structural, functional, or effective—has been crucial for understanding the underlying mechanisms of neural organization. Among these, functional and effective connectivity have received particular attention in computational neuroscience. Functional connectivity reflects statistical dependencies among spatially distinct brain regions, while effective connectivity quantifies causal interactions within neuronal assemblies. Computational approaches to functional connectivity are broadly classified into model-based and model-free methods.

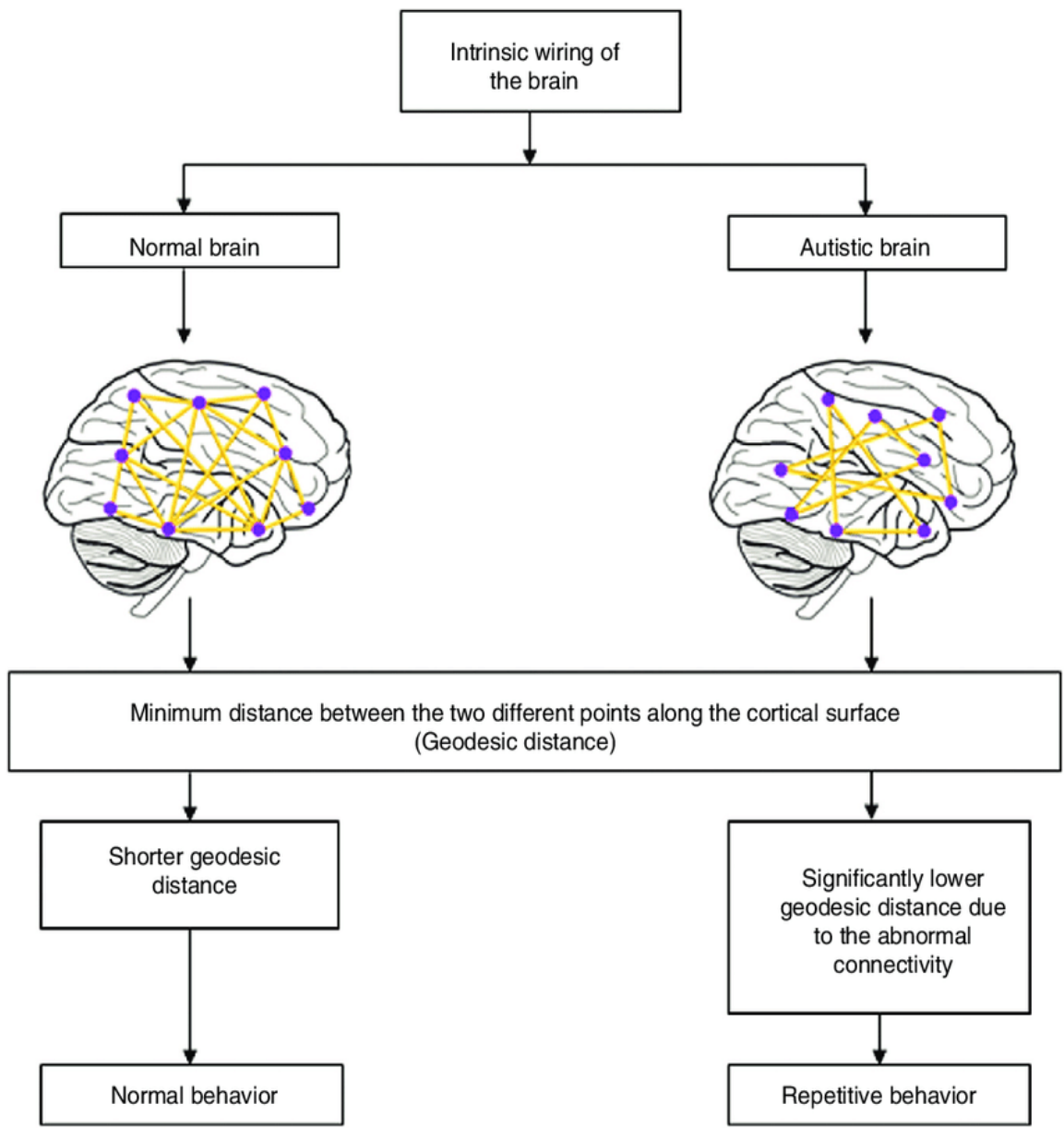

Figure 1: Normal and autistic brain representing connected and disconnected graphs

The human brain exemplifies one of the most complex dynamic systems known. Advances in graph theory have provided a mathematical and conceptual framework to analyze such complexity, offering new opportunities to model and interpret brain networks. Graph theoretical methods model pairwise interactions between neural elements and are widely applied in both functional and effective connectivity research. Figure 1 illustrates the application of graph theory in brain network system that normal brain represents a connected graph whereas autistic brain represents a disconnected graph.[7]

The true potential of graph theory in neuroscience becomes evident after constructing a functional brain network. Various quantitative measures—such as clustering coefficient, modularity, average path length, small-world Ness, assortative, and node centrality—can then be employed to evaluate the topological characteristics of these networks. Through graph-based network analysis, neuroscientists have uncovered key topological characteristics of brain organization, including small- world Ness, modularity, and the existence of highly connected hub regions. A small-world network is characterized by short path lengths between nodes, promoting both local specialization and global integration of information.

Empirical studies have demonstrated that small-world topology varies with cognitive load, developmental stage, and pathological conditions, suggesting its fundamental role in both healthy and diseased states. In this paper we study Cordial labeling and signed product cordial labeling for the brain connectivity network graph and arrive some results interpreted by graph theory.

## 2. Brain Connectivity and Small-World Networks

In neuroscience, the human brain can be represented as a graph $G = (V, E)$ where V denotes distinct cortical or subcortical regions and E represents the structural or functional connections between them. Such networks are typically characterized by small-world properties, combining high clustering with short path lengths, which enable efficient information transfer across the brain [2]. A graph theoritical illustration of brain connectivity network graph (BCNG) is given in Figure 2. this graph has 14 vertices and 18 edges and 3 pendant vertices.

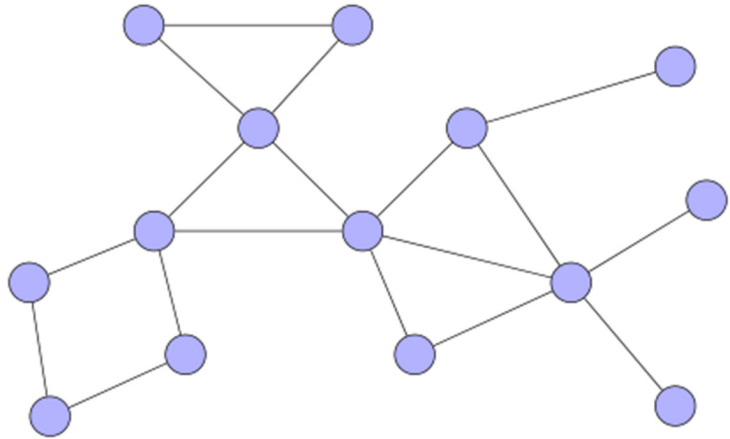

Figure 2: Brain Connectivity network graph-BCNG

A small-world network graph is a type of graph in which most nodes are not directly connected but can be reached from every other node through a small number of steps. It exhibits two key properties — high clustering (nodes tend to form tightly connected groups) and short average path length (any two nodes are connected by relatively few edges). This structure closely mirrors many real-world systems such as social networks, brain connectivity networks, communication systems, and biological networks. The relevance of small-world networks lies in their ability to efficiently balance local specialization and global integration, allowing fast information transfer and robust organization [5]. Application includes neuroscience (understanding brain connectivity), epidemiology (disease spread modeling), computer science (internet and sensor networks), and sociology (analyzing human interaction patterns).

The small-world topology of the human brain persists throughout the entire lifespan [4]. Research has shown that even preterm infants, as early as 30 weeks of gestation, display small-world properties within their brain networks. This organizational structure continues to characterize human neural connectivity from early development through adulthood.

Evolving through natural selection, the small-world configuration of the brain enables efficient segregation and integration of information at minimal cost. This cost–efficiency balance not only optimizes neural communication but also serves as an inspiring model for advancements in brain-like computing and the development of energy-efficient artificial systems [5].

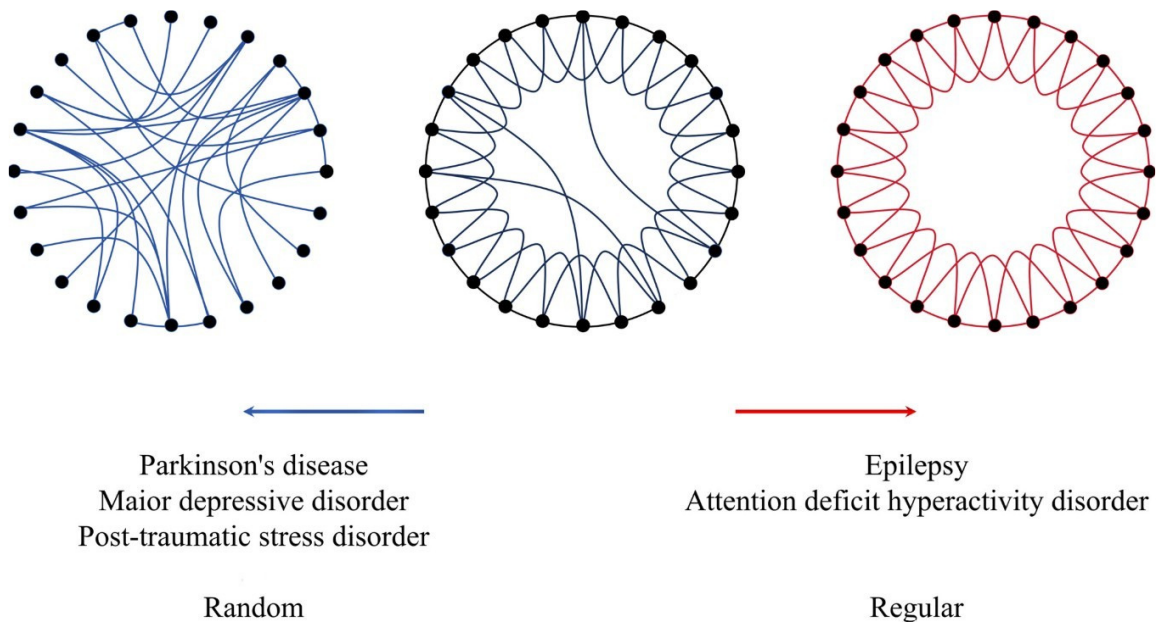

Figure 3: Small world network graph related to brain's connectivity network

In Figure 3, middle graph represents the normal small world network graph of brain network. The left graph represents Parkinson's disease and the right one represents the Epilepsy condition.

## 3. Mathematical Background

### 3.1 Literature on Cordial labeling

In this section, we recall some fundamental concepts and definitions related to graph labeling, which form the basis for our further discussion on cordial and signed product cordial labeling. Graph labeling is a well-studied area in graph theory with numerous variations depending on the mathematical operations used in defining vertex and edge labels. A graph labeling is an assignment of integers to the vertices or edges, or both, subject to certain conditions. The following definitions are adopted for our study.

The concept of cordial labeling was introduced by I. Cahit [3] as a natural relaxation of graceful and harmonious labeling. The motivation behind this concept was to provide a more flexible and inclusive form of labeling that maintains a sense of balance between vertices and edges labeled with 0 and 1.

**Definition 1: (Cordial Labeling)**

A cordial labeling is a binary vertex labeling $f : V(G) \to \{0,1\}$ such that:

1. The number of vertices labeled 0 and 1 differ by at most 1, i.e.,

$$|v_f(0) - v_f(1)| \leq 1$$

2. The induced edge labeling $f^* : E \to \{0,1\}$ is given by

$$f^*(uv) = |f(u) - f(v)|, \text{ such that } |e_f(0) - e_f(1)| \leq 1.$$

If such a labeling exists, the graph G is called a cordial graph. [3]

The concept of Signed Product Cordial Labeling was first introduced by J. Baskar Babujee and L. Shobana [1]. Following this introduction, S. Soundar Rajan and J. Baskar Babujee [6] extended the concept to several classes of graphs.

**Definition 2: (Signed Product Cordial Labeling)**

A vertex labeling of a graph G, $\psi : V(G) \to \{-1,1\}$ with induced edge labeling $\xi : E(G) \to \{-1,1\}$ defined by

$$\xi(uv) = 1 \text{ if } \xi(u) \text{ and } \xi(v) \text{ have same assignment.}$$
$$\xi(uv) = -1 \text{ if } \xi(u) \text{ and } \xi(v) \text{ have different assignment.}$$

is called a signed product cordial labeling (SPCL) if

$$|v_\psi(-1) - v_\psi(1)| \leq 1 \text{ and } |e_\xi(-1) - e_\xi(1)| \leq 1,$$

where $v_\psi(-1)$ is the number of vertices labeled with $-1$, $v_\psi(1)$ is the number of vertices labeled with 1, $e_\xi(-1)$ is the number of edges labeled with $-1$, and $e_\xi(1)$ is the number of edges labeled with 1.

A graph G is signed product cordial if it admits a signed product cordial labeling. [1]

**Example 1:** A Shell graph is created by taking a cycle of 'n' vertices in $C_n$ and adding $n - 3$ chords and all share a common end point called the apex. Figure 4 illustrates the signed product cordial labeling of Shell graph on 6 vertices and 9 edges. 3 vertices are labeled with 1 and 3 vertices are labeled with -1. So the vertex conditions are satisfied. Out of 9 edges, 5 edges are labeled with 1 and 4 edges are labeled with -1. The edge conditions are also satisfied. Hence the Shell graph admits signed product cordial labeling.

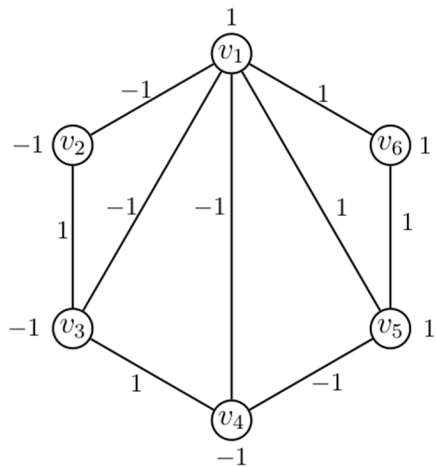

Figure 4: Signed product cordial labeling of Shell graph $S_6$.

**4. Cordial labeling and Brain Connectivity network**

**Theorem 1:** The Brain connectivity network graph admits cordial labeling.

**Proof:**

Let $G = (V, E)$ be the brain-connectivity network graph (Figure 2) with $V = \{v_1, v_2, \ldots, v_{14}\}$ and the edge set $E = \{v_1v_2, v_2v_3, v_3v_4, v_4v_1, v_3v_5, v_6v_7, v_6v_8, v_7v_8, v_6v_3, v_6v_5, v_5v_9, v_9v_{10}, v_9v_{14}, v_5v_{10}, v_5v_{11}, v_{10}v_{11}, v_{10}v_{12}, v_{10}v_{13}\}$.

Define the vertex labeling $f: V \to \{0,1\}$ exactly as shown in the figure by

$$f(v_1) = f(v_2) = f(v_7) = f(v_{11}) = f(v_{12}) = f(v_{13}) = f(v_{14}) = 0$$

The remaining vertices are labeled with 1. The induced edge labeling $f^*: E \to \{0, 1\}$ is given by $f^*(uv) = |f(u) - f(v)|$. We first count vertex labels. The vertices labeled 0 are $V_f(0) = \{v_1, v_2, v_7, v_{11}, v_{12}, v_{13}, v_{14}\}$, hence $|V_f(0)| = 7$.

The vertices labeled with +1 are $V_f(1) = \{v_3, v_4, v_5, v_6, v_8, v_9, v_{10}\}$, so $|V_f(1)| = 7$.

Thus, as literally shown in the figure,

$$|V_f(0)| - |V_f(1)| = |7 - 7| = 0.$$

Next, we compute every induced edge label $f^*(e)$ and count them. For each edge in E, we have:

- $f(v_1v_2) \to 0, f(v_1v_4) \to 1, f(v_2v_3) \to 1, f(v_3v_4) \to 0, f(v_3v_5) \to 0, f(v_3v_6) \to 0,$
- $f(v_5v_6) \to 0, f(v_6v_7) \to 1, f(v_7v_8) \to 1, f(v_8v_9) \to 0, f(v_5v_9) \to 0, f(v_5v_{10}) \to 0,$
- $f(v_9v_{10}) \to 0, f(v_5v_{11}) \to 1, f(v_{11}v_{10}) \to 1, f(v_{10}v_{12}) \to 1, f(v_{10}v_{13}) \to 1,$
  $f(v_9v_{14}) \to 1.$

$$|E_f(0)| = 9 \text{ and } |E_f(1)| = 9. \text{ and}$$

$$|E_f(0) - E_f(1)| = |9 - 9| = 0 \leq 1.$$

Hence, both the vertex and edge conditions are satisfied. Therefore the brain connectivity network graph admits cordial labeling.

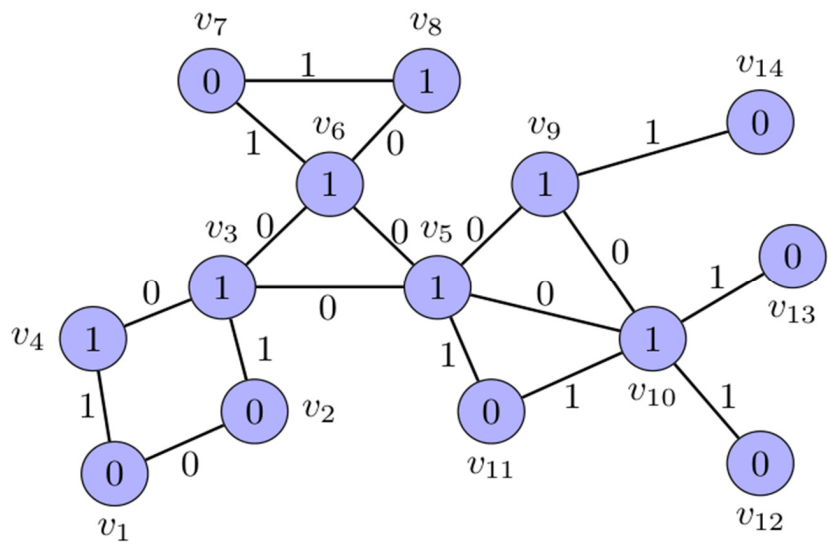

Figure 5: Cordial labeling of Brain connectivity network graph

**Theorem 2:** The brain Connectivity network graph admits signed product cordial labeling.

**Proof:** Let $G = (V, E)$ be the brain-connectivity network graph shown in Figure 2, with $V = \{v_1, v_2, \ldots, v_{14}\}$ and the edge set $E = v_1v_2, v_2v_3, v_3v_4, v_4v_{11}, v_3v_5, v_6v_7,$
$v_7v_8, v6v8, v_5v_9, v_9v_{10}, v_5v_{10}, v_5v_{11}, v_{10}v_{11}, v_{10}v_{12}, v_{10}v_{13}, v_9v_{14}\}.$

Define the vertex labeling $f : V \to \{-1, 1\}$ exactly as shown in the figure by:

- $f(v_1) = f(v_2) = f(v_7) = f(v_{11}) = f(v_{12}) = f(v_{13}) = f(v_{14}) = -1$
- The remaining vertices are labeled with 1.

The induced edge labeling $f^* : E \to \{-1, 1\}$ is given by $f^*(uv) = f(u)f(v)$. We first count vertex labels. The vertices labeled $-1$ are:

$$V_f(-1) = \{v_1, v_2, v_7, v_{11}, v_{12}, v_{13}, v_{14}\} \text{ and } |V_f(-1)| = 7.$$

The remaining vertices are labeled +1 as follows:

$$V_f(1) = \{v_3, v_4, v_5, v_6, v_8, v_9, v_{10}\}.$$

Hence

$$|V_f(-1)| - V_f(1)| = |7 - 7| = 0.$$

Next, we compute every induced edge label $f^*(e)$ and count them. For each edge in $E$ we have:

- $f(v_1v_2) \to 1$, $f(v_1v_4) \to -1$, $f(v_2v_3) \to -1$, $f(v_3v_4) \to 1$, $f(v3v5) \to -1$, $f(v_3v_6) \to 1$,
- $f(v_5v_6) \to -1, f(v_6v_7) \to -1, f(v_7v_8) \to -1, f(v_6v_8) \to -1, f(v_5v_9) \to 1$, $f(v_5v_{10}) \to 1$,
- $f(v_9v_{10}) \to 1, f(v_5v_{11}) \to -1, f(v_{11}v_{10}) \to -1, f(v_{10}v_{12}) \to -1$, $f(v_{10}v_{13}) \to -1$, $f(v_9v_{14}) \to -1$.

$$|E_f(-1)| = 9 \quad \text{and} \quad |E_f(1)| = 18 - 9 = 9.$$

Hence the edge labels are perfectly balanced:

$$|E_f(-1)| - |E_f(1)|| = |9 - 9| = 0 \leq 1.$$

Therefore, the brain connectivity network graph admits signed product cordial labeling.

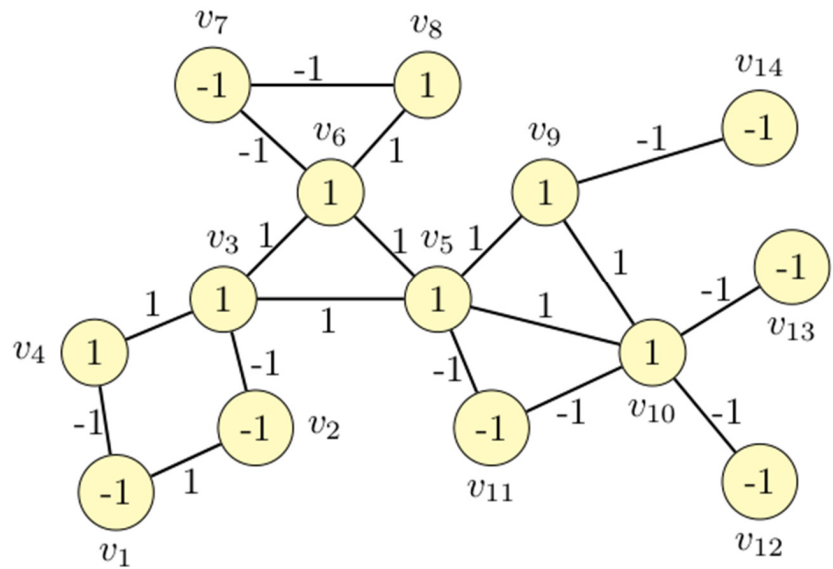

Figure 6: Signed product cordial labeling of brain connectivity network graph

## 5. Results and Discussion

**Observation:** The small world network graph of Parkinson disease do not admit signed product cordial labeling.

Let G = (V, E) be the Small world network graph of Parkinson disease (Figure 3). The Parkinson Disease Graph G has |V| = 24 vertices V = {v₁, v₂, …, v₂₄} and |E| = 28 edges as follows:

$E(G) = \{v_1v_{17}, v_2v_{20}, v_2v_{11}, v_3v_{18}, v_3v_{22}, v_3v_8, v_3v_9, v_3v_{10}, v_4v_8, v_4v_{20},$

$v_5v_{10}, v_5v_{17}, v_6v_{10}, v_7v_{19}, v_9v_{10}, v_{10}v_{20}, v_{11}v_{18}, v_{12}v_{16}, v_{13}v_{18}, v_{14}v_{19},$

$v_{15}v_{24}, v_{16}v_{17}, v_{17}v_{18},\ \ v_{18}v_{19}, v_{18}v_{23}, v_{20}v_{24}, v_{23}v_{24}\}.$

**Case 1:**

Assign $f(v_i) = +1$ if i is odd, and $f(v_i) = -1$ if i is even. Computing the induced edge labels gives:

$$e_f(+1) = 12, \quad e_f(-1) = 16.$$

$$|e_f(+1) - e_f(-1)| = |12 - 16| = 4 > 1.$$

**Case 2:**

Assign $f(v_i) = +1$ for $1 \leq i \leq 12$ and $f(v_i) = -1$ for $13 \leq i \leq 24$. Then:

$$e_f(+1) = 17, \quad e_f(-1) = 11, \text{ and thus}$$

$$|e_f(+1) - e_f(-1)| = |17 - 11| = 6 > 1.$$

Hence, G does not admit signed product cordial labeling under this labeling in either case.

In both cases, the signed product cordiality condition is not satisfied. This finding suggests that when the small-world network graph of the brain fails to admit a signed product cordial labeling, it may be indicative of the presence of Parkinson's disease. These observations imply that the inability of the brain's small-world network graph to admit a signed product cordial labeling may serve as a potential graph-theoretic marker for Parkinson's disease.

The integration of cordial and signed product cordial labeling into neural graph analysis provides a discrete mathematical framework for interpreting structural and functional balance within the brain. This approach opens several theoretical and computational extensions that can be explored in future research.

Brain connectivity is inherently dynamic, with synaptic strengths and functional interactions fluctuating over time. $f_t: V(G_t) \to \{0,1\}$ can be defined on a temporal graph sequence $\{G_t\}_t^T = 1$, where $f_t$ represents the local or global connectivity changes. This approach bridges graph-theoretical formalism with neurobiological interpretation, contributing to the emerging field of mathematical neuroscience. Future work will focus on computational validation using real Neuro-imaging datasets and the derivation of spectral invariants associated with labeling-based balance indices.

## 6. Conclusion

By interpreting the brain connectivity network as a graph structure, we observe that it admits both cordial and signed product cordial labeling. Any deviation from these labeling properties may indicate irregularities or defects in the underlying brain network.

## References

[1] Baskar Babujee. J, L. Shobana, On Signed Product Cordial Labeling, Applied Mathematics, 2, (2011), 1525–1530.

[2] Bullmore. E and O. Sporns, Complex Brain Networks: Graph Theoretical Analysis of Structural and Functional Systems, Nature Reviews Neuroscience, vol. 10, (2009), 186 – 198.

[3] Cahit. I, Cordial Graphs: A Weak Variation of Graceful and Harmonious Graphs, Ars Combinatoria, 23, (1987), 201–207.

[4] Cao M, Wang JH, Dai ZJ, Cao XY, Jiang LL, Fan FM, Song XW, Xia MR, Shu N, Dong Q, Milham MP, Castellanos FX, Zuo XN, He Y, Topological organization of the human brain functional connectome across the lifespan, Dev Cogn Neurosci, 7, (2014), 76-93.

[5] Farahani, Farzad V. and Karwowski, Waldemar and Lighthall, Nichole .R, Application of Graph Theory for Identifying Connectivity Patterns in Human Brain Networks: A Systematic Review, Frontiers in Neuroscience, 13, 2019.

[6] Soundar Rajan.S, J. Baskar Babujee, Further results on signed product cordial labeling, Revista Argentina de Clinica Psicologica, 32, (2023), 1–4.

[7] Watts.D.J and S. H. Strogatz, Collective Dynamics of Small-World Networks, Nature, vol. 393, (1998), pp. 440–442.